%% file: 20ICASSP_beamspars.tex
\documentclass{article}
\usepackage{spconf}
\usepackage{cite}
\usepackage[T1]{fontenc}
\usepackage{graphicx}
\usepackage{amssymb}
\usepackage{amsmath}
\usepackage{paralist}
\usepackage{subfigure}
\usepackage{booktabs} 
\usepackage{algorithm}
\usepackage{algorithmic}
\usepackage{amsthm}
\usepackage{multirow}
\usepackage{microtype} 
\usepackage{tablefootnote}
\usepackage{hyperref}
\usepackage{balance}
\usepackage{breqn}

\input{vmr-symbols-vecbold}
\input{standard-macros}

\input{defs}

\usepackage{color}

\usepackage{enumitem}
\setlist[itemize]{leftmargin=*, itemsep=0.3em, topsep=0.3em} 

\allowdisplaybreaks 

\safemath{\LAMA}{\textrm{LAMA}}
\safemath{\MRT}{\textrm{MRT}}
\safemath{\betamax}{\beta^\text{max}_\setO}
\safemath{\betamaxno}{\beta^\text{max}}
\safemath{\betamin}{\beta^\text{min}_\setO}
\safemath{\betaminno}{\beta^\text{min}}

\safemath{\Nomin}{\No^\textnormal{min}(\beta)}
\safemath{\Nominnobeta}{\No^\text{min}}
\safemath{\Nomax}{\No^\textnormal{max}(\beta)}
\safemath{\Nomaxnobeta}{\No^\textnormal{max}}
\safemath{\EX}{E_\textnormal{x}}
\safemath{\EXP}{\EX^\textnormal{p}}
\safemath{\Eo}{E_0}

\safemath{\tmax}{{t_\textnormal{max}}}
\safemath{\MAP}{\textrm{MAP}}
\safemath{\IO}{\textrm{IO}}
\safemath{\JO}{\textrm{JO}}
\safemath{\Nopost}{N_{0}^\textnormal{post}}
\safemath{\MT}{U}
\safemath{\MR}{B}
\safemath{\Tran}{\textnormal{T}}
\safemath{\Herm}{\textnormal{H}}
\safemath{\row}{\textnormal{r}}
\safemath{\col}{\textnormal{c}}

\safemath{\NT}{N_\textnormal{T}}
\safemath{\DSNR}{\delta \textnormal{SNR}}
\safemath{\betaMOR}{\beta^{\star}}

\title{Sparse Beamspace Equalization for Massive MU-MIMO mmWave  Systems}
\name{Seyed Hadi Mirfarshbafan and Christoph Studer\sthanks{The work of SHM was supported in part by  ComSenTer, one of six centers in JUMP, a SRC program sponsored by DARPA. The work of CS was supported in part by Xilinx, Inc.\ and by the US NSF under grants ECCS-1408006, CCF-1535897,  CCF-1652065, CNS-1717559, and ECCS-1824379.}}
\address{Cornell Tech, New York, NY, 10044;  e-mails:~{sm2675@cornell.edu} and {studer@cornell.edu}}

\begin{document}

\maketitle
	
\input{0-abstract}
%
	
\input{1-introduction}

\input{2-system_model}

\input{3-algorithms}
\input{4-complexity}
\input{5-simulation}
\input{6-conclusion}

\let\oldthebibliography\thebibliography
\let\endoldthebibliography\endthebibliography
\renewenvironment{thebibliography}[1]{
  \begin{oldthebibliography}{#1}
    \setlength{\itemsep}{0.35em}
    \setlength{\parskip}{0em}
}
{
  \end{oldthebibliography}
}
	
\balance
\bibliographystyle{IEEEbib}
\bibliography{bib/VIPabbrv,bib/publishers,bib/confs-jrnls,bib/REFs}
\balance
	
\end{document}

%% file: vmr-symbols-vecbold.tex
\usepackage{amssymb}
\usepackage{amsfonts}
\usepackage{mathrsfs}
\usepackage{xspace}
\usepackage{bm}
\usepackage{upgreek}

\newcommand{\safemath}[2]{\newcommand{#1}{\ensuremath{#2}\xspace}}

\safemath{\bma}{\mathbf{a}}
\safemath{\bmb}{\mathbf{b}}
\safemath{\bmc}{\mathbf{c}}
\safemath{\bmd}{\mathbf{d}}
\safemath{\bme}{\mathbf{e}}
\safemath{\bmf}{\mathbf{f}}
\safemath{\bmg}{\mathbf{g}}
\safemath{\bmh}{\mathbf{h}}
\safemath{\bmi}{\mathbf{i}}
\safemath{\bmj}{\mathbf{j}}
\safemath{\bmk}{\mathbf{k}}
\safemath{\bml}{\mathbf{l}}
\safemath{\bmm}{\mathbf{m}}
\safemath{\bmn}{\mathbf{n}}
\safemath{\bmo}{\mathbf{o}}
\safemath{\bmp}{\mathbf{p}}
\safemath{\bmq}{\mathbf{q}}
\safemath{\bmr}{\mathbf{r}}
\safemath{\bms}{\mathbf{s}}
\safemath{\bmt}{\mathbf{t}}
\safemath{\bmu}{\mathbf{u}}
\safemath{\bmv}{\mathbf{v}}
\safemath{\bmw}{\mathbf{w}}
\safemath{\bmx}{\mathbf{x}}
\safemath{\bmy}{\mathbf{y}}
\safemath{\bmz}{\mathbf{z}}
\safemath{\bmzero}{\mathbf{0}}
\safemath{\bmone}{\mathbf{1}}

\bmdefine{\biad}{a}
\bmdefine{\bibd}{b}
\bmdefine{\bicd}{c}
\bmdefine{\bidd}{d}
\bmdefine{\bied}{e}
\bmdefine{\bifd}{f}
\bmdefine{\bigd}{g}
\bmdefine{\bihd}{h}
\bmdefine{\biid}{i}
\bmdefine{\bijd}{j}
\bmdefine{\bikd}{k}
\bmdefine{\bild}{l}
\bmdefine{\bimd}{m}
\bmdefine{\bind}{n}
\bmdefine{\biod}{o}
\bmdefine{\bipd}{p}
\bmdefine{\biqd}{q}
\bmdefine{\bird}{r}
\bmdefine{\bisd}{s}
\bmdefine{\bitd}{t}
\bmdefine{\biud}{u}
\bmdefine{\bivd}{v}
\bmdefine{\biwd}{w}
\bmdefine{\bixd}{x}
\bmdefine{\biyd}{y}
\bmdefine{\bizd}{z}

\bmdefine{\bixid}{\xi}
\bmdefine{\bilambdad}{\lambda}
\bmdefine{\bimud}{\mu}
\bmdefine{\bithetad}{\theta}
\bmdefine{\biphid}{\phi}
\bmdefine{\bideltad}{\delta}

\safemath{\bmia}{\biad}
\safemath{\bmib}{\bibd}
\safemath{\bmic}{\bicd}
\safemath{\bmid}{\bidd}
\safemath{\bmie}{\bied}
\safemath{\bmif}{\bifd}
\safemath{\bmig}{\bigd}
\safemath{\bmih}{\bihd}
\safemath{\bmii}{\biid}
\safemath{\bmij}{\bijd}
\safemath{\bmik}{\bikd}
\safemath{\bmil}{\bild}
\safemath{\bmim}{\bimd}
\safemath{\bmin}{\bind}
\safemath{\bmio}{\biod}
\safemath{\bmip}{\bipd}
\safemath{\bmiq}{\biqd}
\safemath{\bmir}{\bird}
\safemath{\bmis}{\bisd}
\safemath{\bmit}{\bitd}
\safemath{\bmiu}{\biud}
\safemath{\bmiv}{\bivd}
\safemath{\bmiw}{\biwd}
\safemath{\bmix}{\bixd}
\safemath{\bmiy}{\biyd}
\safemath{\bmiz}{\bizd}

\safemath{\bmxi}{\bixid}
\safemath{\bmlambda}{\bilambdad}
\safemath{\bmmu}{\bimud}
\safemath{\bmtheta}{\bithetad}
\safemath{\bmphi}{\biphid}
\safemath{\bmdelta}{\bideltad}

\safemath{\bA}{\mathbf{A}}
\safemath{\bB}{\mathbf{B}}
\safemath{\bC}{\mathbf{C}}
\safemath{\bD}{\mathbf{D}}
\safemath{\bE}{\mathbf{E}}
\safemath{\bF}{\mathbf{F}}
\safemath{\bG}{\mathbf{G}}
\safemath{\bH}{\mathbf{H}}
\safemath{\bI}{\mathbf{I}}
\safemath{\bJ}{\mathbf{J}}
\safemath{\bK}{\mathbf{K}}
\safemath{\bL}{\mathbf{L}}
\safemath{\bM}{\mathbf{M}}
\safemath{\bN}{\mathbf{N}}
\safemath{\bO}{\mathbf{O}}
\safemath{\bP}{\mathbf{P}}
\safemath{\bQ}{\mathbf{Q}}
\safemath{\bR}{\mathbf{R}}
\safemath{\bS}{\mathbf{S}}
\safemath{\bT}{\mathbf{T}}
\safemath{\bU}{\mathbf{U}}
\safemath{\bV}{\mathbf{V}}
\safemath{\bW}{\mathbf{W}}
\safemath{\bX}{\mathbf{X}}
\safemath{\bY}{\mathbf{Y}}
\safemath{\bZ}{\mathbf{Z}}

\safemath{\bZero}{\mathbf{0}}
\safemath{\bOne}{\mathbf{1}}
\safemath{\bDelta}{\mathbf{\Delta}}
\safemath{\bLambda}{\mathbf{\UpLambda}}
\safemath{\bPhi}{\mathbf{\Upphi}}
\safemath{\bSigma}{\mathbf{\Upsigma}}
\safemath{\bOmega}{\mathbf{\Upomega}}
\safemath{\bTheta}{\mathbf{\Uptheta}}

\bmdefine{\biAd}{A}
\bmdefine{\biBd}{B}
\bmdefine{\biCd}{C}
\bmdefine{\biDd}{D}
\bmdefine{\biEd}{E}
\bmdefine{\biFd}{F}
\bmdefine{\biGd}{G}
\bmdefine{\biHd}{H}
\bmdefine{\biId}{I}
\bmdefine{\biJd}{J}
\bmdefine{\biKd}{K}
\bmdefine{\biLd}{L}
\bmdefine{\biMd}{M}
\bmdefine{\biOd}{N}
\bmdefine{\biPd}{O}
\bmdefine{\biQd}{P}
\bmdefine{\biRd}{R}
\bmdefine{\biSd}{S}
\bmdefine{\biTd}{T}
\bmdefine{\biUd}{U}
\bmdefine{\biVd}{V}
\bmdefine{\biWd}{W}
\bmdefine{\biXd}{X}
\bmdefine{\biYd}{Y}
\bmdefine{\biZd}{Z}

\bmdefine{\biDelta}{\Delta}
\bmdefine{\biLambda}{\Lambda}
\bmdefine{\biPhi}{\Phi}
\bmdefine{\biSigma}{\Sigma}
\bmdefine{\biOmega}{\Omega}
\bmdefine{\biTheta}{\Theta}

\safemath{\bimA}{\biAd}
\safemath{\bimB}{\biBd}
\safemath{\bimC}{\biCd}
\safemath{\bimD}{\biDd}
\safemath{\bimE}{\biEd}
\safemath{\bimF}{\biFd}
\safemath{\bimG}{\biGd}
\safemath{\bimH}{\biHd}
\safemath{\bimI}{\biId}
\safemath{\bimJ}{\biJd}
\safemath{\bimK}{\biKd}
\safemath{\bimL}{\biLd}
\safemath{\bimM}{\biMd}
\safemath{\bimN}{\biNd}
\safemath{\bimO}{\biOd}
\safemath{\bimP}{\biPd}
\safemath{\bimQ}{\biQd}
\safemath{\bimR}{\biRd}
\safemath{\bimS}{\biSd}
\safemath{\bimT}{\biTd}
\safemath{\bimU}{\biUd}
\safemath{\bimV}{\biVd}
\safemath{\bimW}{\biWd}
\safemath{\bimX}{\biXd}
\safemath{\bimY}{\biYd}
\safemath{\bimZ}{\biZd}

\safemath{\bimDelta}{\biDelta}
\safemath{\bimLambda}{\biLambda}
\safemath{\bimPhi}{\biPhi}
\safemath{\bimSigma}{\biSigma}
\safemath{\bimOmega}{\biOmega}
\safemath{\bimTheta}{\biTheta}

\safemath{\setA}{\mathcal{A}}
\safemath{\setB}{\mathcal{B}}
\safemath{\setC}{\mathcal{C}}
\safemath{\setD}{\mathcal{D}}
\safemath{\setE}{\mathcal{E}}
\safemath{\setF}{\mathcal{F}}
\safemath{\setG}{\mathcal{G}}
\safemath{\setH}{\mathcal{H}}
\safemath{\setI}{\mathcal{I}}
\safemath{\setJ}{\mathcal{J}}
\safemath{\setK}{\mathcal{K}}
\safemath{\setL}{\mathcal{L}}
\safemath{\setM}{\mathcal{M}}
\safemath{\setN}{\mathcal{N}}
\safemath{\setO}{\mathcal{O}}
\safemath{\setP}{\mathcal{P}}
\safemath{\setQ}{\mathcal{Q}}
\safemath{\setR}{\mathcal{R}}
\safemath{\setS}{\mathcal{S}}
\safemath{\setT}{\mathcal{T}}
\safemath{\setU}{\mathcal{U}}
\safemath{\setV}{\mathcal{V}}
\safemath{\setW}{\mathcal{W}}
\safemath{\setX}{\mathcal{X}}
\safemath{\setY}{\mathcal{Y}}
\safemath{\setZ}{\mathcal{Z}}
\safemath{\emptySet}{\varnothing}

\safemath{\colA}{\mathscr{A}}
\safemath{\colB}{\mathscr{B}}
\safemath{\colC}{\mathscr{C}}
\safemath{\colD}{\mathscr{D}}
\safemath{\colE}{\mathscr{E}}
\safemath{\colF}{\mathscr{F}}
\safemath{\colG}{\mathscr{G}}
\safemath{\colH}{\mathscr{H}}
\safemath{\colI}{\mathscr{I}}
\safemath{\colJ}{\mathscr{J}}
\safemath{\colK}{\mathscr{K}}
\safemath{\colL}{\mathscr{L}}
\safemath{\colM}{\mathscr{M}}
\safemath{\colN}{\mathscr{N}}
\safemath{\colO}{\mathscr{O}}
\safemath{\colP}{\mathscr{P}}
\safemath{\colQ}{\mathscr{Q}}
\safemath{\colR}{\mathscr{R}}
\safemath{\colS}{\mathscr{S}}
\safemath{\colT}{\mathscr{T}}
\safemath{\colU}{\mathscr{U}}
\safemath{\colV}{\mathscr{V}}
\safemath{\colW}{\mathscr{W}}
\safemath{\colX}{\mathscr{X}}
\safemath{\colY}{\mathscr{Y}}
\safemath{\colZ}{\mathscr{Z}}

\safemath{\opA}{\mathbb{A}}
\safemath{\opB}{\mathbb{B}}
\safemath{\opC}{\mathbb{C}}
\safemath{\opD}{\mathbb{D}}
\safemath{\opE}{\mathbb{E}}
\safemath{\opF}{\mathbb{F}}
\safemath{\opG}{\mathbb{G}}
\safemath{\opH}{\mathbb{H}}
\safemath{\opI}{\mathbb{I}}
\safemath{\opJ}{\mathbb{J}}
\safemath{\opK}{\mathbb{K}}
\safemath{\opL}{\mathbb{L}}
\safemath{\opM}{\mathbb{M}}
\safemath{\opN}{\mathbb{N}}
\safemath{\opO}{\mathbb{O}}
\safemath{\opP}{\mathbb{P}}
\safemath{\opQ}{\mathbb{Q}}
\safemath{\opR}{\mathbb{R}}
\safemath{\opS}{\mathbb{S}}
\safemath{\opT}{\mathbb{T}}
\safemath{\opU}{\mathbb{U}}
\safemath{\opV}{\mathbb{V}}
\safemath{\opW}{\mathbb{W}}
\safemath{\opX}{\mathbb{X}}
\safemath{\opY}{\mathbb{Y}}
\safemath{\opZ}{\mathbb{Z}}
\safemath{\opZero}{\mathbb{O}}
\safemath{\identityop}{\opI}

\safemath{\veca}{\bma}
\safemath{\vecb}{\bmb}
\safemath{\vecc}{\bmc}
\safemath{\vecd}{\bmd}
\safemath{\vece}{\bme}
\safemath{\vecf}{\bmf}
\safemath{\vecg}{\bmg}
\safemath{\vech}{\bmh}
\safemath{\veci}{\bmi}
\safemath{\vecj}{\bmj}
\safemath{\veck}{\bmk}
\safemath{\vecl}{\bml}
\safemath{\vecm}{\bmm}
\safemath{\vecn}{\bmn}
\safemath{\veco}{\bmo}
\safemath{\vecp}{\bmp}
\safemath{\vecq}{\bmq}
\safemath{\vecr}{\bmr}
\safemath{\vecs}{\bms}
\safemath{\vect}{\bmt}
\safemath{\vecu}{\bmu}
\safemath{\vecv}{\bmv}
\safemath{\vecw}{\bmw}
\safemath{\vecx}{\bmx}
\safemath{\vecy}{\bmy}
\safemath{\vecz}{\bmz}

\safemath{\veczero}{\bmzero}
\safemath{\vecone}{\bmone}
\safemath{\vecxi}{\bmxi}
\safemath{\veclambda}{\bmlambda}
\safemath{\vecmu}{\bmmu}
\safemath{\vectheta}{\bmtheta}
\safemath{\vecphi}{\bmphi}
\safemath{\vecdelta}{\bmdelta}

\safemath{\matA}{\bA}
\safemath{\matB}{\bB}
\safemath{\matC}{\bC}
\safemath{\matD}{\bD}
\safemath{\matE}{\bE}
\safemath{\matF}{\bF}
\safemath{\matG}{\bG}
\safemath{\matH}{\bH}
\safemath{\matI}{\bI}
\safemath{\matJ}{\bJ}
\safemath{\matK}{\bK}
\safemath{\matL}{\bL}
\safemath{\matM}{\bM}
\safemath{\matN}{\bN}
\safemath{\matO}{\bO}
\safemath{\matP}{\bP}
\safemath{\matQ}{\bQ}
\safemath{\matR}{\bR}
\safemath{\matS}{\bS}
\safemath{\matT}{\bT}
\safemath{\matU}{\bU}
\safemath{\matV}{\bV}
\safemath{\matW}{\bW}
\safemath{\matX}{\bX}
\safemath{\matY}{\bY}
\safemath{\matZ}{\bZ}
\safemath{\matzero}{\bmzero}

\safemath{\matDelta}{\bDelta}
\safemath{\matLambda}{\bLambda}
\safemath{\matPhi}{\bPhi}
\safemath{\matSigma}{\bSigma}
\safemath{\matOmega}{\bOmega}
\safemath{\matTheta}{\bTheta}

\safemath{\matidentity}{\matI}
\safemath{\matone}{\matO}

\safemath{\rnda}{A}
\safemath{\rndb}{B}
\safemath{\rndc}{C}
\safemath{\rndd}{D}
\safemath{\rnde}{E}
\safemath{\rndf}{F}
\safemath{\rndg}{G}
\safemath{\rndh}{H}
\safemath{\rndi}{I}
\safemath{\rndj}{J}
\safemath{\rndk}{K}
\safemath{\rndl}{L}
\safemath{\rndm}{M}
\safemath{\rndn}{N}
\safemath{\rndo}{O}
\safemath{\rndp}{P}
\safemath{\rndq}{Q}
\safemath{\rndr}{R}
\safemath{\rnds}{S}
\safemath{\rndt}{T}
\safemath{\rndu}{U}
\safemath{\rndv}{V}
\safemath{\rndw}{W}
\safemath{\rndx}{X}
\safemath{\rndy}{Y}
\safemath{\rndz}{Z}

\safemath{\rveca}{\bimA}
\safemath{\rvecb}{\bimB}
\safemath{\rvecc}{\bimC}
\safemath{\rvecd}{\bimD}
\safemath{\rvece}{\bimE}
\safemath{\rvecf}{\bimF}
\safemath{\rvecg}{\bimG}
\safemath{\rvech}{\bimH}
\safemath{\rveci}{\bimI}
\safemath{\rvecj}{\bimJ}
\safemath{\rveck}{\bimK}
\safemath{\rvecl}{\bimL}
\safemath{\rvecm}{\bimM}
\safemath{\rvecn}{\bimN}
\safemath{\rveco}{\bomO}
\safemath{\rvecp}{\bimP}
\safemath{\rvecq}{\bimQ}
\safemath{\rvecr}{\bimR}
\safemath{\rvecs}{\bimS}
\safemath{\rvect}{\bimT}
\safemath{\rvecu}{\bimU}
\safemath{\rvecv}{\bimV}
\safemath{\rvecw}{\bimW}
\safemath{\rvecx}{\bimX}
\safemath{\rvecy}{\bimY}
\safemath{\rvecz}{\bimZ}

\safemath{\rvecxi}{\bmxi}
\safemath{\rveclambda}{\bmlambda}
\safemath{\rvecmu}{\bmmu}
\safemath{\rvectheta}{\bmtheta}
\safemath{\rvecphi}{\bmphi}

\safemath{\rmatA}{\bimA}
\safemath{\rmatB}{\bimB}
\safemath{\rmatC}{\bimC}
\safemath{\rmatD}{\bimD}
\safemath{\rmatE}{\bimE}
\safemath{\rmatF}{\bimF}
\safemath{\rmatG}{\bimG}
\safemath{\rmatH}{\bimH}
\safemath{\rmatI}{\bimI}
\safemath{\rmatJ}{\bimJ}
\safemath{\rmatK}{\bimK}
\safemath{\rmatL}{\bimL}
\safemath{\rmatM}{\bimM}
\safemath{\rmatN}{\bimN}
\safemath{\rmatO}{\bimO}
\safemath{\rmatP}{\bimP}
\safemath{\rmatQ}{\bimQ}
\safemath{\rmatR}{\bimR}
\safemath{\rmatS}{\bimS}
\safemath{\rmatT}{\bimT}
\safemath{\rmatU}{\bimU}
\safemath{\rmatV}{\bimV}
\safemath{\rmatW}{\bimW}
\safemath{\rmatX}{\bimX}
\safemath{\rmatY}{\bimY}
\safemath{\rmatZ}{\bimZ}

\safemath{\rmatDelta}{\bimDelta}
\safemath{\rmatLambda}{\bimLambda}
\safemath{\rmatPhi}{\bimPhi}
\safemath{\rmatSigma}{\bimSigma}
\safemath{\rmatOmega}{\bimOmega}
\safemath{\rmatTheta}{\bimTheta}

%% file: standard-macros.tex
\usepackage{amssymb}
\usepackage{amsfonts}
\usepackage{mathrsfs}
\usepackage{xspace}
\usepackage{bm}
\usepackage{fancyref}
\usepackage{textcomp}

\usepackage{multirow}
\usepackage{stmaryrd}

\newenvironment{textbmatrix}{	\setlength{\arraycolsep}{2.5pt}%
								\big[\begin{matrix}}{\end{matrix}\big]%
								\raisebox{0.08ex}{\vphantom{M}}}

\def\be{\begin{equation}}
\def\ee{\end{equation}}
\def\een{\nonumber \end{equation}}
\def\mat{\begin{bmatrix}}
\def\emat{\end{bmatrix}}
\def\btm{\begin{textbmatrix}}
\def\etm{\end{textbmatrix}}

\def\ba#1\ea{\begin{align}#1\end{align}}
\def\bas#1\eas{\begin{align*}#1\end{align*}}
\def\bs#1\es{\begin{split}#1\end{split}}
\def\bg#1\eg{\begin{gather}#1\end{gather}}
\def\bml#1\eml{\begin{multline}#1\end{multline}}
\def\bi#1\ei{\begin{itemize}#1\end{itemize}}

\newcommand{\lefto}{\mathopen{}\left}

\DeclareMathOperator*{\argmin}{arg\;min}		
\DeclareMathOperator*{\argmax}{arg\;max}		
\DeclareMathOperator{\Exop}{\opE}			

\newcommand{\Ex}[2]{\ensuremath{\Exop_{#1}\lefto[#2\right]}} 	

\safemath{\dirac}{\delta}					
\safemath{\krond}{\dirac}					

\safemath{\upto}{\uparrow}
\safemath{\downto}{\downarrow}
\safemath{\iu}{j}							
\safemath{\ev}{\lambda}						
\safemath{\hilseqspace}{l^{2}}				
\newcommand{\banachfunspace}[1]{\setL^{#1}}	
\safemath{\hilfunspace}{\banachfunspace{2}}	

\safemath{\SNR}{\textit{SNR}} 				
\safemath{\PAR}{\textit{PAR}} 				
\safemath{\No}{N_0}							
\safemath{\Es}{E_s}							
\safemath{\Eb}{E_b}							
\safemath{\EbNo}{\frac{\Eb}{\No}}
\safemath{\EsNo}{\frac{\Es}{\No}}

\DeclareMathOperator{\CHop}{\ensuremath{\opH}} 
\safemath{\tvir}{\rndh_{\CHop}}				
\safemath{\tvtf}{\rndl_{\CHop}}				
\safemath{\spf}{\rnds_{\CHop}}				
\safemath{\bff}{H_{\CHop}}					

\safemath{\ircf}{r_{h}}						
\safemath{\tftvcf}{r_{s}}					
\safemath{\tfcf}{r_{l}}						
\safemath{\bfcf}{r_{H}}						

\safemath{\tcorr}{c_h}						
\safemath{\scf}{c_{s}}						
\safemath{\tfcorr}{c_{l}}					
\safemath{\fcorr}{c_{H}}						

\safemath{\mi}{I}							
\safemath{\capacity}{C}						

\safemath{\normal}{\mathcal{N}}			
\safemath{\jpg}{\mathcal{CN}}			
\safemath{\mchain}{\leftrightarrow}		

\safemath{\dB}{\,\mathrm{dB}}
\safemath{\dBm}{\,\mathrm{dBm}}
\safemath{\Hz}{\,\mathrm{Hz}}
\safemath{\kHz}{\,\mathrm{kHz}}
\safemath{\MHz}{\,\mathrm{MHz}}
\safemath{\GHz}{\,\mathrm{GHz}}
\safemath{\s}{\,\mathrm{s}}
\safemath{\ms}{\,\mathrm{ms}}
\safemath{\mus}{\,\mathrm{\text{\textmu}s}}
\safemath{\ns}{\,\mathrm{ns}}
\safemath{\ps}{\,\mathrm{ps}}
\safemath{\meter}{\,\mathrm{m}}
\safemath{\mm}{\,\mathrm{mm}}
\safemath{\cm}{\,\mathrm{cm}}
\safemath{\m}{\,\mathrm{m}}
\safemath{\W}{\,\mathrm{W}}
\safemath{\mW}{\, \mathrm{mW}}
\safemath{\J}{\,\mathrm{J}}
\safemath{\K}{\,\mathrm{K}}
\safemath{\bit}{\,\mathrm{bit}}
\safemath{\nat}{\,\mathrm{nat}}

\safemath{\define}{\triangleq}			

\safemath{\equivalent}{\sim}
\safemath{\distas}{\sim}					
\safemath{\sdiff}{\Delta}				

\safemath{\reals}{\mathbb{R}}
\safemath{\positivereals}{\reals_{+}}
\safemath{\integers}{\mathbb{Z}}
\safemath{\posint}{\integers_{+}}
\safemath{\naturals}{\mathbb{N}}
\safemath{\posnaturals}{\naturals_{+}}
\safemath{\complexset}{\mathbb{C}}
\safemath{\rationals}{\mathbb{Q}}

\newcommand*{\fancyrefapplabelprefix}{app}		
\newcommand*{\fancyrefthmlabelprefix}{thm}		
\newcommand*{\fancyreflemlabelprefix}{lem}		
\newcommand*{\fancyrefcorlabelprefix}{cor}		
\newcommand*{\fancyrefdeflabelprefix}{def}		
\newcommand*{\fancyrefproplabelprefix}{prop}		
\newcommand*{\fancyrefexmpllabelprefix}{exmpl}
\newcommand*{\fancyrefalglabelprefix}{alg}		
\newcommand*{\fancyreftbllabelprefix}{tbl}		

\frefformat{vario}{\fancyrefseclabelprefix}{Section~#1}
\frefformat{vario}{\fancyrefthmlabelprefix}{Theorem~#1}
\frefformat{vario}{\fancyreftbllabelprefix}{Table~#1}
\frefformat{vario}{\fancyreflemlabelprefix}{Lemma~#1}
\frefformat{vario}{\fancyrefcorlabelprefix}{Corollary~#1}
\frefformat{vario}{\fancyrefdeflabelprefix}{Definition~#1}
\frefformat{vario}{\fancyreffiglabelprefix}{Fig.~#1}
\frefformat{vario}{\fancyrefapplabelprefix}{Appendix~#1}
\frefformat{vario}{\fancyrefeqlabelprefix}{(#1)}
\frefformat{vario}{\fancyrefproplabelprefix}{Proposition~#1}
\frefformat{vario}{\fancyrefexmpllabelprefix}{Example~#1}
\frefformat{vario}{\fancyrefalglabelprefix}{Algorithm~#1}

%% file: defs.tex
\safemath{\dictab}{[\,\dicta\,\,\dictb\,]}

\safemath{\ysig}{\bmy}
\safemath{\ysighat}{\hat{\ysig}}
\safemath{\ysigdim}{M}
\safemath{\xsig}{\bmx}
\safemath{\xsigdim}{N}
\safemath{\nx}{n_x}
\safemath{\zsig}{\bmz}
\safemath{\zsigdim}{\ysigdim}
\safemath{\rsig}{\bmr}
\safemath{\Adict}{\bA}
\safemath{\Adicttilde}{\widetilde{\Adict}}
\safemath{\Adictdim}{\outputdim\times\xsigdim}
\safemath{\avec}{\bma}
\safemath{\avectilde}{\tilde{\avec}}
\safemath{\Bdict}{\bB}
\safemath{\Bdicttilde}{\widetilde{\Bdict}}
\safemath{\Cdict}{\bC}
\safemath{\cvec}{\bmc}
\safemath{\Ddict}{\bD}
\safemath{\Ddictdim}{\ysigdim\times\xsigdim}
\safemath{\dvec}{\bmd}
\safemath{\Ddicttilde}{\widetilde{\bD}}
\safemath{\Bonb}{\bB}
\safemath{\bvec}{\bmb}
\safemath{\Bonbdim}{\ysigdim\times\ysigdim}
\safemath{\noise}{\bmn}
\safemath{\noisedim}{\ysigim}
\safemath{\err}{\bme}
\safemath{\errdim}{\ysigdim}
\safemath{\errset}{\setE}
\safemath{\nerr}{n_e}
\safemath{\delop}{\bP_\errset}
\safemath{\delopc}{\bP_{{\errset}^c}}

\safemath{\cplxi}{\imath}
\safemath{\cplxj}{\jmath}

\safemath{\dict}{\matD}
\safemath{\inputdim}{N}		
\safemath{\outputdim}{M}		
\safemath{\sparsity}{S}	
\safemath{\inputdimA}{{N_a}}	
\safemath{\inputdimB}{{N_b}}	
\safemath{\elemA}{{n_a}}	
\safemath{\elemB}{{n_b}}	
\safemath{\resA}{\matR_a}	
\safemath{\resB}{\matR_b}	
\safemath{\subD}{\matS} 
\safemath{\subA}{\matS_a} 
\safemath{\subB}{\matS_b} 
\safemath{\dicta}{\matA} 	
\safemath{\dictb}{\matB} 	
\safemath{\hollowS}{H}
\safemath{\hollowA}{H_a}
\safemath{\hollowB}{H_b}
\safemath{\cross}{Z}
\safemath{\coh}{\mu_d}			
\safemath{\coha}{\mu_a}			
\safemath{\cohb}{\mu_b}			
\safemath{\mubs}{\nu}	
\safemath{\cohm}{\mu_m} 
\safemath{\dictset}{\setD}	
\safemath{\dictsetp}{\dictset(\coh,\coha,\cohb)}	
\safemath{\dictsetgen}{\dictset_\text{gen}}
\safemath{\dictsetgenp}{\dictsetgen(\coh)}
\safemath{\dictsetonb}{\dictset_\text{onb}}
\safemath{\dictsetonbp}{\dictsetonb(\coh)}

\safemath{\leftside}{U}
\safemath{\rightsideA}{R_a}
\safemath{\rightsideB}{R_b}

\safemath{\indexS}{\setI_S} 

\safemath{\na}{n_a}			
\safemath{\nb}{n_b}			
\safemath{\coeffa}{p_i}	
\safemath{\coeffb}{q_j}	
\safemath{\seta}{\setP}		
\safemath{\setb}{\setQ}     
\safemath{\setw}{\setW}	
\safemath{\setz}{\setZ}	
\safemath{\cola}{\veca}		
\safemath{\colb}{\vecb}		
\safemath{\cold}{\vecd}		
\safemath{\inputvec}{\vecx} 	
\safemath{\error}{\vece}	
\safemath{\noiseout}{\vecz} 	
\safemath{\inputvecel}{x}
\safemath{\inputveca}{\vecx_a}
\safemath{\inputvecb}{\vecx_b}
\safemath{\outputvec}{\vecy}	
\safemath{\lambdamin}{\lambda_{\mathrm{min}}}

\safemath{\elltwo}{\ell_2}
\safemath{\ellone}{\ell_1}
\safemath{\ellzero}{\ell_0}
\safemath{\ellinf}{\ell_\infty}
\safemath{\ellinftilde}{\ell_{\widetilde\infty}}
\safemath{\licard}{Z(\coh,\coha,\cohb)}
\safemath{\xsol}{\hat{x}}
\safemath{\xbord}{x_b}		
\safemath{\xstat}{x_s}		
\safemath{\xstatLone}{\tilde{x}_s}
\safemath{\order}{\mathcal{O}} 
\safemath{\scales}{\Theta} 
\safemath{\ones}{\mathbf{1}} 
\safemath{\zeroes}{\mathbf{0}} 
\safemath{\thlone}{\kappa(\coh,\cohb)} 
\safemath{\constoneA}{\delta} 
\safemath{\constoneB}{\epsilon} 
\safemath{\nlarge}{L}				   
\safemath{\sumlarge}{S_\nlarge}
\safemath{\maxlarger}{P_\nlarge}	   
\safemath{\Pzero}{\textrm{P0}}	
\safemath{\Pone}{\textrm{P1}}
\safemath{\vecfir}{\vecw}			 
\safemath{\vecsec}{\vecz}
\safemath{\elvecfir}{w}              
\safemath{\elvecsec}{z}				 
\safemath{\nlargefir}{n}
\safemath{\normout}{\gamma}
\safemath{\auxfun}{h}
\safemath{\supp}{\textrm{supp}}

\safemath{\indexa}{\ell}
\safemath{\indexb}{r}
\safemath{\indexc}{i}
\safemath{\indexd}{j}

\safemath{\project}{P}

%% file: 0-abstract.tex

\begin{abstract}

We propose equalization-based data detection algorithms for all-digital millimeter-wave (mmWave) massive multiuser multiple-input multiple-out (MU-MIMO) systems that exploit sparsity in the beamspace domain to reduce complexity. We provide a condition on the number of users, basestation antennas, and channel sparsity for which beamspace equalization can be less complex than conventional antenna-domain processing. We evaluate the performance-complexity trade-offs of existing and new beamspace equalization algorithms using simulations with realistic mmWave channel models. Our results reveal that one of our proposed beamspace equalization algorithms achieves up to $8 \times$ complexity reduction under line-of-sight conditions, assuming a sufficiently large number of transmissions within the channel coherence interval.

\end{abstract}


%% file: 1-introduction.tex

\section{Introduction} 
\label{sec:intro}

Millimeter-wave (mmWave) communication~\cite{mmWillWork,mmWaveWireless} and massive multiuser multiple-input  multiple-output (MU-MIMO)~\cite{LarssonMMIMONextGen} are key technologies of next-generation wireless  systems.
The large portions of unused mmWave frequency bands promise significantly increased data rates but also require higher sampling rates, which complicates analog and digital hardware design.
Since wave propagation at mmWave frequencies is directional and real-world channels typically comprise only a small number of dominant propagation paths~\cite{mmWillWork,mmWaveWireless}, the channel vectors associated with each user in the beamspace domain are sparse~\cite{BEACHESSPAWC, AlkhateebChEstHybrid2014,schniter2014channel,deng2018mmwave, AlkhateebLimitedFeedback, LeeOMP}. 
Therefore, a promising approach to reduce complexity of massive MU-MIMO mmWave systems is to perform baseband processing in the beamspace domain. 

Beamspace processing to reduce complexity has been proposed for single-user mmWave MIMO systems with hybrid analog-digital front-ends in \cite{CAPMIMO, Song13Beamspace}. The case of MU-MIMO systems has been studied in \cite{SayeedGLOBECOM, GaoNearOptimalBeamspace}, where beam-selection is used to reduce the dimension of the processing tasks by exploiting channel sparsity in the beamspace domain.
%
Beamspace processing in all-digital basestation architectures has gained recent attention in \cite{dumbLund,mahdavi2018low, localLMMSE}. The equalizer proposed in \cite{localLMMSE}, called \textit{Local LMMSE}, identifies a contiguous block of beams with a mean-squared error (MSE) criterion. The papers \cite{dumbLund,mahdavi2018low} propose low-complexity beam-selection algorithms and corresponding hardware designs; we refer to the method in \cite{dumbLund} as the \textit{strongest beams} (SB) algorithm.
%

\noindent \textit{\textbf{Contributions:}} We propose new sparsity-exploiting equalization algorithms and identify conditions for which beamspace processing is able to reduce the complexity compared to conventional antenna-domain processing. We investigate the performance-complexity trade-offs of beamspace-domain equalization with all-digital basestation architectures. 
%

%
\noindent \textit{\textbf{Notation:}} Boldface lowercase and uppercase letters represent column vectors and matrices, respectively. For a matrix $\bA$, the transpose and Hermitian transpose is  $\bA^\Tran$ and $\bA^\Herm$, respectively, and the $k$th column is $\bma_k = [\bA]_k$.
For a vector $\bma$, the  $k$th entry is $a_k$ = $[\bma]_k$.
The column vector~$\bma^r_m$ is the transpose of the $m$th row of matrix $\bA$.
The $\ell_2$-norm of $\bma$ is $\|\veca\|$; the Frobenius norm of $\bA$ is $\|\bA\|_F$.
The $N\times N$ identity and $N\times N$ discrete Fourier transform (DFT) matrices are denoted by $\bI_N$ and~$\bF$, respectively; the DFT matrix satisfies $\bF\bF^H=\bI_N$.
%
%

%% file: 2-system_model.tex

\section{Beamspace MIMO System Model}
\label{sec:systemmodel}

%
%
%
%
We consider a mmWave massive MU-MIMO  uplink system in which~$U$ single-antenna user equipments (UEs) transmit data in the same time-frequency resource to a basestation (BS) equipped with a $B$-antenna uniform linear array (ULA).
We focus on  frequency-flat and block-fading channels for which the channel is assumed to stay constant over a block of $T$ channel uses. Let $\bar{\bH} \in \complexset^{B\times U}$ denote the channel matrix for a given coherence time interval. The \emph{antenna domain} received signal vector at the BS is modeled as %
$\bar{\bmy} = \bar{\bH}\bms + \bar{\bmn}$,
%
where the vector $\bms\in\setS^U$ contains the data symbols transmitted by all UEs, with the power constraint $\Ex{}{|s_u|^2} = E_s$, $u=1,\ldots,U$.
The vector $\bar{\bmn} \sim \setC\setN(\bZero_{B\times1},N_0 \bI_B)$ models thermal noise with variance $\No$. By applying the DFT to the received antenna-domain vector $\bar{\bmy}$, we obtain the \emph{beamspace} input-output relation
%
$\bmy=\bF\bar{\bmy} = \bH\bms + \bmn$.
%
Here, the matrix $\bH = \bF \bar{\bH}$ and vector $\bmn=\bF \bar{\bmn}$ are the beamspace channel matrix and  beamspace noise vector, respectively. 

\subsection{Equalization-Based Data Detection}
\label{sec:lindetection}
Equalization-based data detection in the antenna domain typically consists of two phases: \emph{preprocessing} and \emph{equalization}. 
During preprocessing, which is only performed once per coherence interval, an equalization matrix $\bar{\bW}$ is computed based on an estimate of the channel matrix.
During equalization, which is performed for every received vector ($T$ times per coherence interval), estimates  $\hat{\bms} = \bar{\bW}\bar{\bmy}$ of the transmit vector are generated. 
Beamspace equalization resembles the above procedure with the exception that the equalization matrix $\bW$ is computed from the estimated beamspace-domain channel matrix $\hat{\bH}$, and is applied to the beamspace receive vectors~$\bmy$. We will focus on computation of equalization matrices in \fref{sec:algorithms}.

Since wave propagation at mmWave frequencies is directional~\cite{akdeniz2014millimeter,RappaportmmWaveModel}, the beamspace channel vectors $\bmh_u$ associated with each UE $u=1,\ldots,U$, that correspond to columns of~$\bH$, are approximately sparse \cite{schniter2014channel}, i.e., most of the channel's energy is concentrated on a few incident angles. 
Each row $b = 1, \ldots, B$  of the beamspace domain channel matrix $\bH$, and each entry of the received vector $\bmy$, correspond to one spatial angle-of-arrival (AoA); the indices~$b$ are referred to as \emph{beam indices}.
The sparsity in beamspace domain offers the opportunity reduce the complexity of baseband processing, including that of equalization-based data detection.
%

%% file: 3-algorithms.tex

\section{Sparse Beamspace Equalization} \label{sec:algorithms}

Linear minimum mean-square error (LMMSE) equalization is among the most prominent data-detection methods. The beamspace LMMSE equalization matrix~$\bW$ is computed as
\begin{align} \label{eq:LMMSE}
\bW =\argmin_{\tilde{\bW}\in\complexset^{U\times B}} \| \bI_U - \tilde{\bW} \bH\|_F^2+\rho \| \tilde{\bW} \|_F^2,
\end{align} 
where $\rho = N_0/E_s$. A closed-form solution to \fref{eq:LMMSE} is given by $\bW = (\bH^\Herm \bH + \rho \bI)^{-1}\bH^\Herm$.
In what follows, we propose and investigate algorithms that exploit the sparsity of mmWave channels in the beamspace domain with the goal of computing equalization matrices $\hat{\bW}$ with fewer nonzero elements than the full matrix $\bW$. Such sparse beamspace equalization matrices reduce the number of multiplications required for each equalization task, which has the potential to decrease hardware complexity and power dissipation in all-digital BS architectures. 
Sparsity-exploiting equalization algorithms require an input parameter~$\delta\in[0,1]$, referred to as the \emph{density coefficient}, which describes a beamspace equalization matrix~$\hat{\bW}$ with only $\delta BU$ nonzero entries. 
The density coefficient $\delta$ is an input parameter to the preprocessing algorithm and affects the error rate depending on the actual channel sparsity. Sparsity-exploiting algorithms can be categorized into  (i) column-wise and (ii) entry-wise methods. Column-wise methods select a subset of beam indices $\{1, \dots, B\}$ as the support set to construct an equalization matrix with only $K = \delta B$ nonzero columns. Entry-wise methods select the support set of each row of $\hat{\bW}$ independently from other rows and construct an equalization matrix with $K = \delta B$ nonzero entries per row.

\subsection{Columnwise Orthogonal Matching Pursuit (COMP)}
\label{sec:COMP}
We start by proposing a column-wise orthogonal matching pursuit (COMP) algorithm, which tries to find a solution $\hat{\bW}$ for the MSE criterion in \fref{eq:LMMSE} that only consists of $K = \delta B$ nonzero columns---the remaining columns are zero. 
COMP performs $K$ iterations and successively identifies one of the $K$ nonzero columns of~$\hat{\bW}$ in each iteration in a greedy fashion. 

Let us define $\Omega^{(k)}$ as the support set consisting of the indices of $k$ nonzero columns that COMP has selected during iterations $1,\ldots,k$. Initially, we set $\Omega^{(0)} = \varnothing$. 
We use $\hat{\bW}^{(k)}$ to denote the $U \times k$ matrix computed after the  $k$th iteration, and $\bH_{\Omega^{(k)}} = \bH(\Omega^{(k)}, :)$ to represent the $k \times U$ matrix containing the rows of $\bH$ indexed by the set $\Omega^{(k)}$. 
Each COMP iteration consists of two steps:

\noindent {\bf\em Step 1) Select Beam Index:} Assuming that $k$ beam indices are collected in $\Omega^{(k)}$ during iterations $1$ to $k$, COMP identifies the $(k+1)$th best beam index $b^{(k+1)}$, by solving  
\begin{align} \label{eq:COMP1}
b^{(k+1)} =\!\!\!\!\! \argmin_{{b^\prime} \in \{1,\ldots,B\}\textbackslash \Omega^{(k)}}  \min_{\tilde{\bmw}\in\complexset^{U}}  \| \bA^{(k)}  \!-\! \tilde{\bmw} ({\bmh}_{b^\prime}^r)^{\Tran}\|_F^2 
+\rho \| \tilde{\bmw} \|^2,
\end{align}
where $({\bmh}_{b^\prime}^r)^{\Tran}$ is the $b^\prime$th row of $\bH$, and $\bA^{(k)}$ is defined as
\begin{align} \label{eq:updateA}
\bA^{(k)} = \bI_U -\hat{\bW}^{(k)} \bH_{\Omega^{(k)}},
\end{align} 
with initialization $\bA^{(0)} = \bI_U$.
The solution to \fref{eq:COMP1} is given by 
\begin{align} \label{eq:column_index}
b^{(k+1)} =\argmax_{{b^\prime} \in \{1,\ldots,B\}\textbackslash\Omega^{(k)}}  \frac{\| \bA^{(k)} \bmh^r_{b^\prime}\|^2}{\| \bmh_{b^\prime}^r \|^2 + \rho}.
\end{align} 
With $b^{(k+1)}$, we update the support $\Omega^{(k+1)} = \Omega^{(k)} \cup b^{(k+1)}$.

\noindent {\bf\em Step 2) Compute Equalization Matrix $\hat{\bW}^{(k+1)}$:}
The MSE-optimal equalization matrix $\hat{\bW}^{(k+1)}$ with only $k+1$ columns determined by the support set $\Omega^{(k+1)}$, is given by
\begin{align} \label{eq:lmmse_COMP}
\hat{\bW}^{(k+1)} =(\bH^\Herm_{\Omega^{(k+1)}}\bH_{\Omega^{(k+1)}}+ \rho \bI_U)^{-1} \bH^\Herm_{\Omega^{(k+1)}}.
\end{align} 

We note that, since in each iteration only one element is added to $\Omega^{(k)}$ to form $\Omega^{(k+1)}$, the matrix $(\bH^\Herm_{\Omega^{(k+1)}}\bH_{\Omega^{(k+1)}}+ \rho \bI_U)$ in \fref{eq:lmmse_COMP} is a rank-one update to the matrix from the previous iteration. Hence, we use the Sherman-Morrison formula \cite{NumRecipes} to avoid an explicit matrix inversion in iterations $2$ to $K$ to reduce complexity. 
%
In summary, each COMP iteration consists of computing \fref{eq:updateA}, \fref{eq:column_index}, and \fref{eq:lmmse_COMP}. After $K$ iterations, the output of COMP is the sparse equalization matrix $\hat{\bW}^{(K)} \in \complexset^{U \times K}$.
\subsection{Largest Columns (LC) Approximation}

As an approximate, low-complexity alternative to COMP, the support set $\Omega$ can be populated by simply collecting $K$ beam indices
that maximize the objective function in \fref{eq:column_index} with $\bA^{(k)} = \bI_U$, which corresponds to the rows of  $\bH$ with the largest $\ell_2$ norms. Then, the $U \times K$ equalization matrix is constructed as in \fref{eq:lmmse_COMP}. The resulting method has been proposed in \cite{mahdavi2018low}, and we call it the \textit{largest columns} (LC) approximation.
\subsection{Entrywise Orthogonal Matching Pursuit (EOMP)}
In contrast to COMP, EOMP constructs the support set for each row of $\hat{\bW}$   independently from the other rows. 
%
Mathematically, the optimization problem \fref{eq:LMMSE} can be decomposed into~$U$ independent problems
\begin{align} \label{eq:rowLMMSE}
\hat{\bmw}^r_u =\argmin_{\tilde{\bmw}^r_u \in \complexset^{B}}  \| \bme_u - \bH^\Tran \tilde{\bmw}^r_u \|^2 + \rho \| \tilde{\bmw}^r_u \|^2 ,
\end{align}
where $\bme_u$ is the $u$th column of $\bI_U$, and~$\tilde{\bmw}^r_u$ denotes a column vector equal to the transpose of the $u$th row of $\tilde{\bW}$.
For each $u = 1, \ldots, U$, EOMP finds a solution to \fref{eq:rowLMMSE} with the constraint that only $K = \delta B$ entries of~$\hat{\bmw}^r_u$ are nonzero. To this end, for each of the~$U$ rows of $\hat{\bW}$, EOMP performs $K$ iterations similar to those of COMP.
We denote the $k$-element support set for the $u$th row of $\hat{\bW}$ obtained during iterations $1$ to $k$ by $\Omega_u^{(k)}$. Initially, we set $\Omega_u^{(0)} = \varnothing$. We also use $\hat{\bmw}^{r(k)}_u$ to denote the $k$-entry vector that is computed at the end of the $k$th iteration for the $u$th row of $\hat{\bW}$. 
Each EOMP iteration for the $u$th row of $\hat{\bW}$ consists of two steps:

\noindent {\bf\em Step 1) Select Beam Index:} 
%
EOMP selects the $(k+1)$th beam index, by solving
\begin{align} \label{eq:rowLMMSE1}
b^{(k+1)} = \!\!\!\!\!  \argmin_{{b^\prime} \in \{1,\ldots,B\} \textbackslash \Omega_u^{(k)}} \min_{\tilde{w} \in \complexset} \,  \| \bmz^{(k)} - \tilde{w} \bmh^r_{b^\prime} \|^2 + \rho | \tilde{w}|^2,
\end{align} 
where the residual vector $\bmz^{(k)}$ is given by
\begin{align} \label{eq:updatez}
\bmz^{(k)} = \bme_u -\bH^\Tran_{\Omega^{(k)}} \hat{\bmw}^{r(k)}_u,
\end{align} 
with initialization $\bmz^{(0)} = \bme_u$.
The solution to \fref{eq:rowLMMSE1} is given by the following expression:
\begin{align} \label{eq:EOMP_entry_index}
b^{(k+1)} =\argmax_{{b^\prime} \in \{1,\ldots,B\} \textbackslash \Omega_u^{(k)}} \frac{| (\bmz^{(k)})^\Herm \bmh^r_{b^\prime} |^2}{\| \bmh_{b^\prime}^r \|^2 + \rho}.
\end{align} 
With $b^{(k+1)}$, we update the support as $\Omega_u^{(k+1)} = \Omega_u^{(k)} \cup b^{(k+1)}$.

\noindent {\bf\em Step 2) Compute Equalization Vector $\hat{\bmw}^{r(k+1)}_u$:}
Given the new support set $\Omega_u^{(k+1)}$, the optimal $(k+1)$-entry vector for the $u$-th row of the equalization matrix is computed as
\begin{align} \label{eq:lmmse_EOMP}
\hat{\bmw}^{r(k+1)}_u = \bH_{\Omega^{(k+1)}} (\bH^\Herm_{\Omega^{(k+1)}}\bH_{\Omega^{(k+1)}}+ \rho \bI_U)^{-1}\bme_u.
\end{align} 

As in \fref{sec:COMP}, we use the Sherman-Morrison formula~\cite{NumRecipes} to avoid an explicit matrix inversion in \fref{eq:lmmse_EOMP}.
In summary, each iteration of EOMP consists of computing~\fref{eq:updatez}, \fref{eq:EOMP_entry_index}, and \fref{eq:lmmse_EOMP}. After $K$ iterations, the output of the algorithm $\hat{\bmw}^{r(K)}_u$, is a $K$-dimensional vector that contains the nonzero entries of the $u$th row of the equalization matrix $\hat{\bW}$. This procedure is applied for each row of $\hat{\bW}$, independently.
\subsection{Largest Entries (LE) Approximation}
As an approximate, low-complexity alternative to EOMP, the support set $\Omega_u$ for the $u$th row of $\hat{\bW}$ is obtained by gathering the top-$K$
indices of \fref{eq:EOMP_entry_index} with $\bmz^{(k)} = \bme_u$, which corresponds to the entries of $\hat{\bmw}^r_u$ with the largest absolute values. Then, the nonzero entries indexed by~$\Omega_u$ are computed according to \fref{eq:lmmse_EOMP}. This procedure is carried out for each row of $\hat{\bW}$ independently, and we refer to it as the \textit{largest entries} (LE) approximation. 

%

%% file: 4-complexity.tex

\section{Complexity Analysis} \label{sec:complexity}

We now provide a complexity analysis for the proposed sparsity-exploiting equalization algorithms and for existing methods in terms of the number of real-valued multiplications required during preprocessing and equalization.
%
\fref{tbl:complexity} summarizes the real-valued multiplications for each equalization algorithm; we assume that each complex multiplication requires four real-valued multiplications. The quantity $E = 4TUK$ corresponds to the complexity required for applying the $U \times K$ beamspace equalization matrix to $T$ received vectors within a coherence time. 
The quantity $F = (U+T)(2B \log_2 B)$ corresponds to the number of multiplications required by the fast Fourier transform \cite{SplitRadixFFT}, applied to $U$ columns of the channel matrix and $T$ received antenna-domain vectors. 
All algorithms involve computations of the form $(\bH_{\Omega}^\Herm \bH_{\Omega} + \rho \bI)^{-1}\bH_{\Omega}^\Herm$, which requires $2U^3 + 6KU^2 - (2K+1)U$ multiplications for a $K \times U$ matrix~$\bH_{\Omega}$ by taking into account symmetries and using the Cholesky decomposition for matrix inversion. For the local LMMSE method in \cite{localLMMSE}, we use the procedure put forward in \cite[Sec.~III.A-3]{localLMMSE} to minimize complexity. 

We observe in \fref{tbl:complexity}, that the computational complexity of all sparsity-exploiting methods decreases for smaller density factors $K = \delta B$. However, smaller density factors typically incur a higher performance loss. The associated complexity-performance trade-offs are investigated in \fref{sec:simres}. 
\begin{table}[tp]
	\centering
	\renewcommand{\arraystretch}{1.1}
	\begin{minipage}[c]{1\columnwidth}
		\centering
		\caption{Complexity of sparsity-exploiting algorithms.}
		\vspace{0.15cm}
		\label{tbl:complexity}
		\resizebox{0.85\columnwidth}{!}{
			\begin{tabular}{@{}ll@{}}
				\toprule
				Algorithm & Number of real-valued multiplications \\
				\midrule
				LMMSE  & $2U^3 + 6BU^2 - 2(B+1)U + 4TUB $ \\
				\hline
				Local LMMSE  & $((-4U-6)K^3 + (4BU+8B+2U)K^2$\\
				\cite{localLMMSE} & $(8BU-12B+4U-6)K)+ E+ F$  \\
				\hline
				SB \cite{dumbLund} & $2BU + 2U^3 + 6KU^2 - 2(K+1)U+ E + F$\\
				\hline
				COMP & $2U^3 + (4BK + 2K^2 + 12K - 4)U^2 + $\\
				& $(2B+2BK-2K^2+4K-6)U+ E + F$\\	
				\hline
				LC \cite{mahdavi2018low} & $6BU + 2U^3 + 6KU^2 - 2KU - 2U+ E + F$ \\	
				\hline
				EOMP & $2U^4 +(6K-4)U^3 + (3K^2+(2B+9)K)U^2$ \\
				& $+(2B(K+1)-K^2)U+ E + F$ \\
				\hline
				LE & $2U^4 + 2KU^3 + (4K-2)U^2 + 2BU+ E + F$\\
				\bottomrule
		\end{tabular}}
	\end{minipage}
\end{table}
\subsection{Asymptotic Complexity Analysis}
We now analyze the asymptotic complexity of sparsity-exploiting equalization algorithms when the coherence time~$T$ approaches infinity, i.e., where the preprocessing complexity becomes irrelevant. 
%
%
For antenna-domain processing, estimating the transmit symbol of each UE involves $s_u = (\bar{\bmw}^r_u)^\Tran \bar{\bmy}$, $u = 1, \ldots, U$, which corresponds to $4UB$ real-valued multiplications. 
Beamspace equalization requires one FFT for each received vector, corresponding to approximately $2B \log_2 B$ real-valued multiplications \cite{SplitRadixFFT}. Column-wise methods with a $U \times K$ equalization matrix and entry-wise methods with $K$-entry equalization vectors per UE require $4UK$ multiplications for each equalization task. Thus, a necessary condition for the beamspace equalization to have lower complexity than the antenna-domain equalization is $4UB > 4UK + 2 B \log_2 B$, which is equivalent to
$\delta < 1 -  \frac{\log_2 B}{2U}$,
where $\delta = K/B$ is the density coefficient. This expression reveals two conditions for beamspace equalization to be less complex than the antenna-domain equalization: 
(i) Since $\delta > 0$, we must have $U > \frac{1}{2} \log_2 B$ and (ii) the ratio of selected beams out of $B$ total beams must be smaller than $1 - \frac{1}{2}(\log_2 B)/U$. We note that this asymptotic analysis  provides only a necessary condition and does not take into account preprocessing. 

%% file: 5-simulation.tex

\section{Performance-Complexity Trade-offs}
\label{sec:simres}

\begin{figure}[tp]
	\centering
	\subfigure[LoS, $B=128$, $U=16$]{\includegraphics[width=0.495\columnwidth]{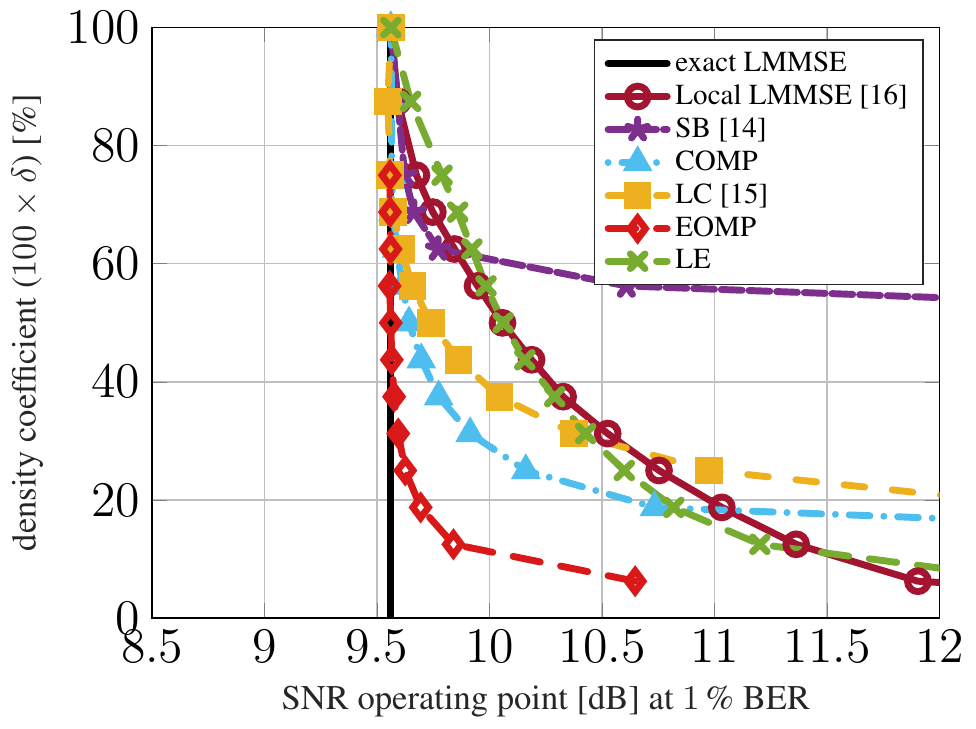}}
	\subfigure[Non-LoS, $B=128$, $U=16$]{\includegraphics[width=0.495\columnwidth]{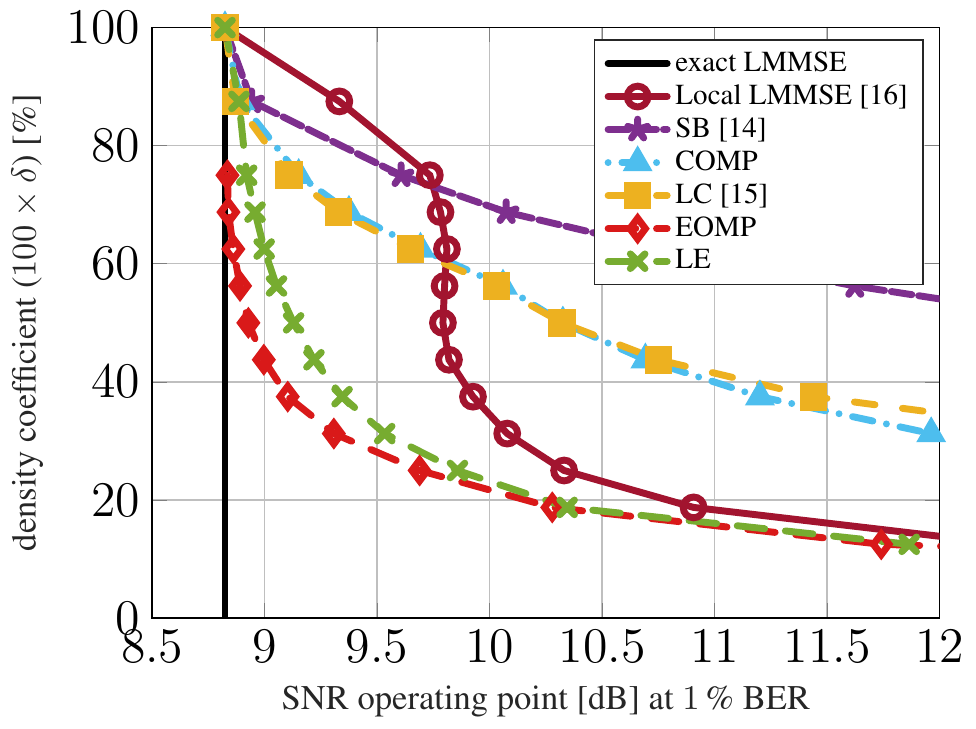}}
	\vspace{-0.4cm}
	 \caption{SNR operating point at 1\% uncoded BER.}
	\label{fig:SNRop}
	\vspace{-0.2cm}
\end{figure}

\begin{figure}[tp]
	\centering
	\subfigure[LoS, $B=128$, $U=16$]{\includegraphics[width=0.495\columnwidth]{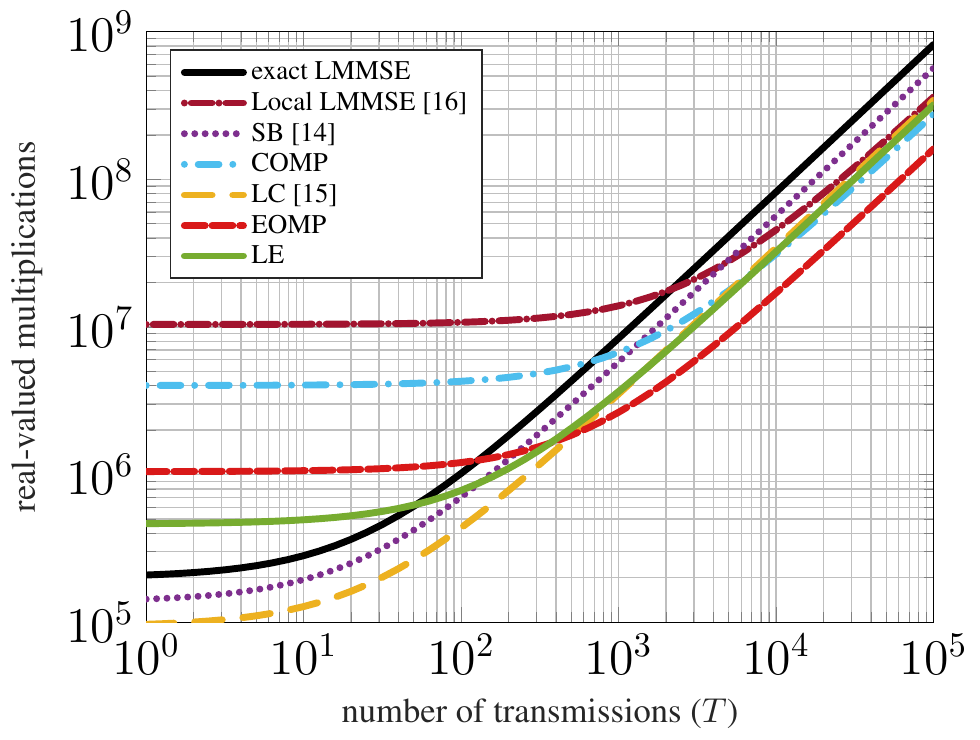}}
	\subfigure[Non-LoS, $B=128$, $U=16$]{\includegraphics[width=0.495\columnwidth]{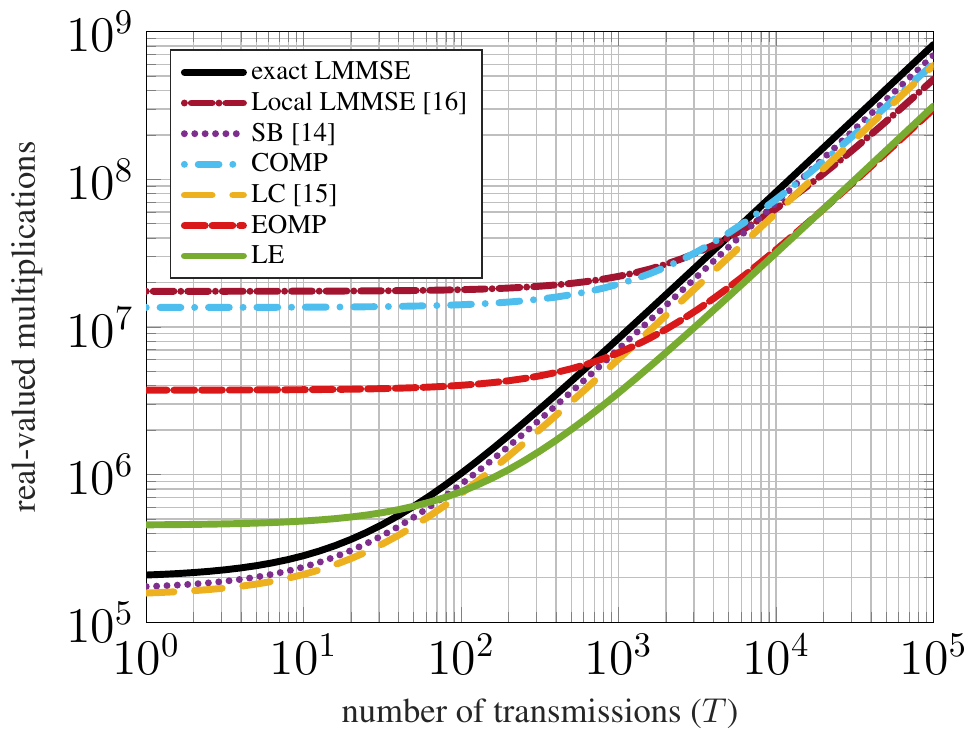}}
	\vspace{-0.4cm}
	\caption{Complexity determined by the density coefficient ($\delta_{\text{min}}$) required to achieve an SNR gap of at most $1$\,dB with respect to the exact LMMSE at 1\% uncoded BER.}
	\label{fig:comp}
\end{figure}

We now evaluate the performance-complexity trade-offs of the proposed and existing sparsity-exploiting algorithms. 
%
We simulate a massive MU-MIMO system with $B = 128$ BS antennas and $U=16$ single antenna UEs transmitting $16$-QAM symbols.
The channel matrices for both line-of-sight (LoS) and non-LoS scenarios are generated using the QuaDRiGa mmMAGIC UMi model~\cite{QuaDRiGa} at a carrier frequency of 60\,GHz with a ULA using $\lambda/2$ antenna spacing.
UEs are placed randomly in a $120^\circ$ circular sector with minimum and maximum distance of $10$\,m and $110$\,m from the BS antenna array, respectively, and with at least $1^\circ$ angular separation between UEs.
We use pilot-based channel estimation and BEACHES~\cite{BEACHESSPAWC} to denoise the channel estimates in the beamspace domain.

%
\noindent \textit{\textbf{SNR Operating Point:}}
To evaluate the performance of sparsity-exploiting algorithms under different density parameters, we simulate uncoded bit error rate (BER) versus SNR, for each algorithm for a range of density coefficients $\delta$ from $0.03$ to $1$.
\fref{fig:SNRop} shows the SNR operating point of each algorithm to achieve $\textit{BER}=10^{-2}$ for both LoS (\fref{fig:SNRop}(a)) and non-LoS (\fref{fig:SNRop}(b)) channel conditions. The exact antenna-domain LMMSE algorithm is represented by a vertical line, as the density coefficient does not apply to this algorithm.
%
%
We see from \fref{fig:SNRop} that for each sparsity-exploiting algorithm, lower density parameters (smaller values of $\delta$) require higher SNR operating points to achieve the same BER. We also observe that the proposed EOMP algorithm outperforms all other equalization algorithms for LoS and non-LoS channels.

%
\noindent \textit{\textbf{Performance vs.\ Complexity:}}
Due to the disparity between the performance and complexity of sparsity-exploiting beamspace equalization algorithms, a unified comparison approach is necessary to gain insight into the required complexity (in terms of the multiplication count) of each algorithm, without incurring a significant performance loss.
For each algorithm we identify the minimum density coefficient $\delta_{\text{min}}$ that results in no more than $1$\,dB SNR gap with respect to  the exact LMMSE at 1\% uncoded BER.
\fref{fig:comp} shows the number of multiplications from \fref{tbl:complexity} for each algorithm corresponding to $K =\delta_{\text{min}} B$, versus number of transmissions $T$, for both LoS (\fref{fig:comp}(a)) and non-Los (\fref{fig:comp}(b)) channels. The number of transmissions $T$ within a channel coherence interval is proportional to the product of the coherence time $T_c$ and the communications bandwidth $BW$. Therefore, for a mmWave channel with $BW=500$ MHz and $T_c = 1$\,ms \cite{RailwayMMWave}, the number of coherent transmissions $T$ can be up to $10^{5}$. We observe in \fref{fig:comp} that the complexity savings of sparsity-exploiting equalization manifests\ itself mainly for large values of $T$. 
In this regime (i.e., for $T > 10^{4}$), the proposed EOMP algorithm achieves the lowest complexity (due to small $\delta_{\text{min}}$) in both LoS and non-LoS scenarios with $6 \times$ to $8 \times$ complexity reduction compared to antenna-domain LMMSE equalization.

%% file: 6-conclusion.tex
\section{Conclusions}

We have proposed three novel  beamspace equalization algorithms that leverage  angular sparsity of mmWave propagation to reduce complexity. 
Our simulations have shown that the proposed EOMP algorithm is able to outperform existing methods both in terms of required SNR operating point and complexity. 
In addition, our asymptotic complexity analysis has revealed two necessary conditions for sparsity-exploiting beamspace equalization to be less complex than antenna-domain equalization: (i)  the number of UEs $U$ must be at least $\frac{1}{2} \log_2 B$ and (ii) the density coefficient $\delta$ must be below $1 - \frac{1}{2}\log_2 B/U$. 
In addition, our investigation pinpoints three ingredients of successful beamspace-domain processing: (i) systems with long coherence time, (ii) low-complexity beamspace transforms (FFTs), and (iii) low-complexity preprocesssing algorithms.